\documentclass[3p,times,procedia]{elsarticle}
\usepackage{nupha_ecrc}

\volume{00}

\firstpage{1}

\journalname{Nuclear Physics A}

\runauth{}

\jid{nupha}

\jnltitlelogo{Nuclear Physics A}

\usepackage{amssymb}

\usepackage[figuresright]{rotating}

\begin{document}

\begin{frontmatter}



\dochead{XXVIIIth International Conference on Ultrarelativistic Nucleus-Nucleus Collisions\\ (Quark Matter 2019)}

\title{$p_T$-dependent multiplicity fluctuations from PCA and initial  conditions}


\author[UNIFAL,SACLAY]{Fernando G. Gardim}
\author[USP]{Fr\'ed\'erique Grassi}
\author[USP]{Pedro Ishida}
\author[USP]{Matthew Luzum}
\author[SACLAY]{Jean-Yves Ollitrault}
\address[UNIFAL]{Instituto de Ci\^encia e Tecnologia, Universidade Federal de Alfenas,
  37715-400 Po\c cos de Caldas, MG, Brazil
}
\address[USP]{
    Instituto de F\'{i}sica, Universidade de S\~{a}o Paulo, 
    R. do Mat\~{a}o 1371, 05508-090 S\~{a}o Paulo, SP, Brazil
}
\address[SACLAY]{
    Institut de physique th\'eorique, Universit\'e Paris Saclay, CNRS, CEA,
    F-91191 Gif-sur-Yvette, France
}

\begin{abstract}
  We present a  Principal Component Analysis  for a hydrodynamic simulation  and compare with CMS experimental data. While the results are reasonable for anisotropic flow, for  multiplicity fluctuations they are qualitatively different.
  We argue that this is due to too large transverse momentum ($p_T$) fluctuations and $N-p_T$ covariance in the simulation than seen experimentally.
  In turn this is related to too large initial size fluctuations.
\end{abstract}


\end{frontmatter}


\section{Introduction}
The importance of initial state fluctuations to understand flow observables is now well established. 
The Principal Component Analysis (PCA) is used in many fields of science and  tailored to study fluctuations. So in \cite{Bhalerao:2014mua}, it was suggested to use it to study event-by-event fluctuations in relativistic nuclear collisions. The advantage of the method
compared to more traditional ones is that it incorporates all the information in
two-particle azimuthal correlations  in a single setting.

To carry on the PCA, the single particle distribution is expanded as
\begin{equation}
    \frac{dN}{dy dp_T d\phi}    =  \sum_{n=-\infty}^{+\infty} \mathcal V_n(p_T)e^{-in\phi},
\end{equation}
where $\phi$ is the azimuthal angle $p_T$. Then the covariance matrix is constructed, diagonalized and written in term of its eigenvectors (starting from the largest eigenvalue):
\begin{equation}
  \mathcal V_{n\Delta}(p_T^a,p_T^b)  \equiv 
    \langle \mathcal V_n(p_T^a)\mathcal V_n^*(p_T^b) \rangle -
    \langle \mathcal V_n(p_T^a) \rangle \langle \mathcal V_n^*(p_T^b) \rangle 
 =   \sum_{\alpha} \mathcal V_n^{(\alpha)}(p_T^a) \mathcal V_n^{(\alpha)}(p_T^b),
\label{vnd}
\end{equation}
where the average is performed over events, and $\mathcal V_n^{(\alpha)}(p_T)$ is the eigenvector. The principal components can be scaled in order to obtain the standard flow  

\begin{equation}
v_n^{(\alpha)}(p_T) \equiv   \frac{\mathcal V_n^{(\alpha)}(p_T)}{\langle \mathcal V_0(p_T) \rangle}.
\end{equation}

First experimental results on PCA have been presented by the CMS collaboration \cite{Sirunyan:2017gyb}. The physical meaning of the principal components was discussed in \cite{Bhalerao:2014mua,Mazeliauskas:2015vea,Mazeliauskas:2015efa,Liu:2019jxg} and further developments were obtained in \cite{Bozek:2017thv,Hippert:2019swu}.  In \cite{Gardim:2019iah}, we argued that hydrodynamic simulations are in agreement with CMS data for anisotropic flow  (n=2,3) but as far as we know, none reproduces, even qualitatively,   multiplicity fluctuations ($n=0$). We then  investigated the origin of this discrepancy.  In this contribution, we summarize the $n=0$ PCA case and present some new development.

\section{Results for multiplicity fluctuations}

\begin{figure*}
  \centering
\includegraphics[width=0.8\textwidth]{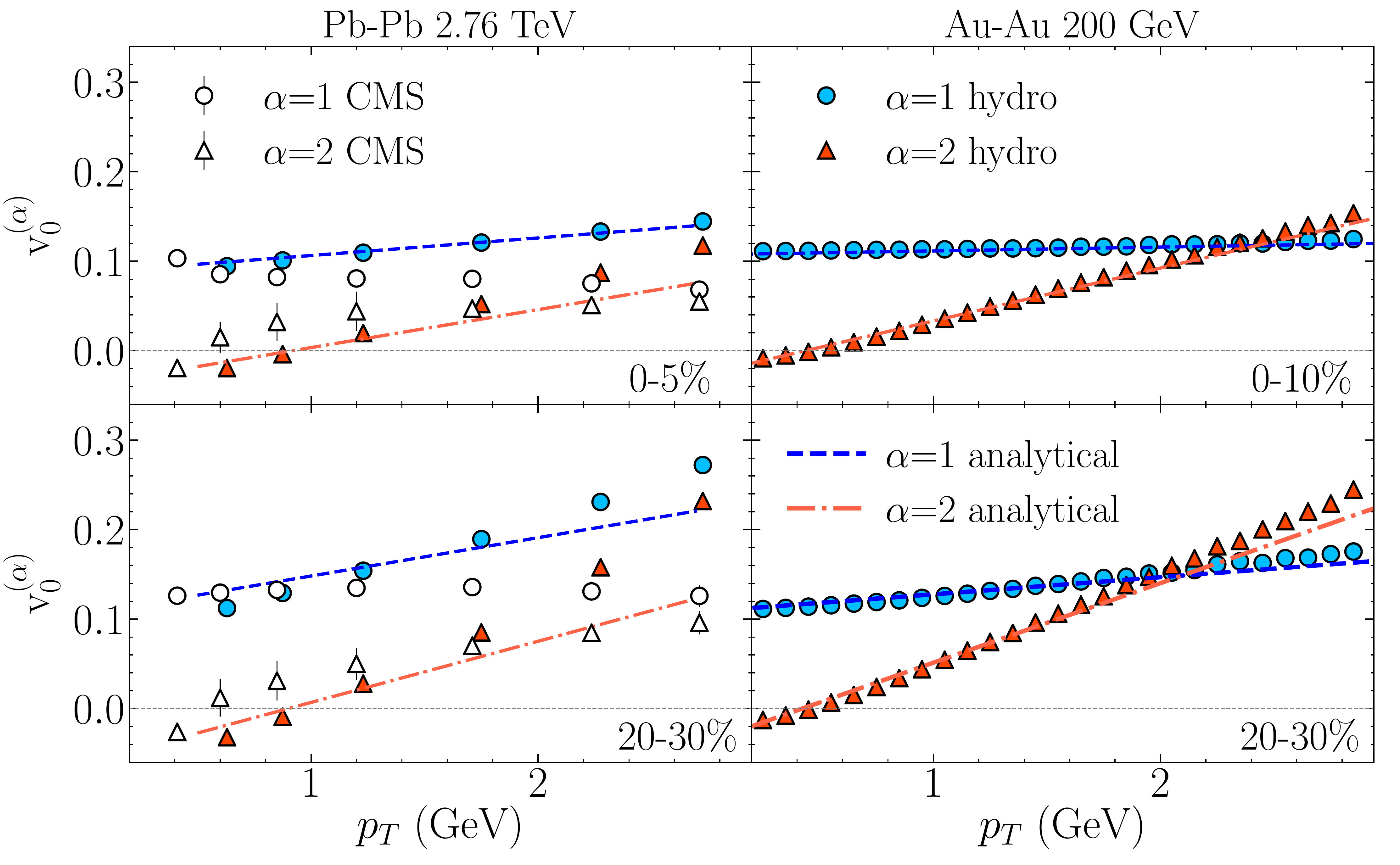} 
\caption{
First two scaled principal components for $n=0$ (multiplicity fluctuations). For LHC, a  comparison between our ideal fluid calculation, CMS data~\cite{Sirunyan:2017gyb} and the approximate result from the toy model, eq.~(\ref{eq:pptoy}) (lines) is shown. For RHIC, our predictions from the hydrodynamic simulation and the analytical model are presented.
}
\label{fig:fig1QM} 
\end{figure*}

A comparison between our hydrodynamics results, coming from NeXSPheRIO 3+1D perfect ideal fluid \cite{Gardim:2019iah}, and CMS data as well as predictions for RHIC are displayed in figure~\ref{fig:fig1QM}. At LHC, the leading component of eq. (\ref{vnd}), i.e the largest eigenvalue, is rather independent of $p_T$ in experiment, while it increases with $p_T$ in our hydrodynamic calculation. The increase is less strong at RHIC energies. The increase at LHC energies is not specific to our implementation, as it has been seen by other groups~\cite{Mazeliauskas:2015efa, Matt}.

To pinpoint the origin of this discrepancy, an analytical model can be built. Assuming the $p_T$ spectrum as
\begin{equation}
    \label{spectra}
    \frac{1}{2\pi}\frac{dN}{dydp_T} =
    \mathcal V_0(p_T) =
    \frac{2 p_T N}{\pi{\bar{p}_T^2}} e^{-\frac{2 p_T}{\bar{p}_T}},
\end{equation}
we can let multiplicity $N$ deviate from mean multiplicity $\langle N \rangle$ as $ N = \langle N \rangle + \delta N$, and momentum from average momentum as $ \bar p_T  = \langle \bar p_T \rangle + \delta \bar p_T$. Using eq.~(\ref{spectra}) on the covariance matrix, eq.~(\ref{vnd}), the eigenvectors can be computed analytically when $\mathcal V_{n\Delta}(p_T^a,p_T^b)$ is diagonalized. The first two components, leading $\alpha=1$ and subleading $\alpha=2$, are 
\begin{equation}
v_0^{(1)}(p_T)  \simeq 
    \frac{\sigma_N}{\langle N \rangle} +
    \left[\frac{
        - \left(\frac{\sigma_{p_T}}{\langle \bar{p}_T \rangle}\right)^2 +
          2 \frac{
              \langle \delta N \delta\bar p_T \rangle
          }{
              \langle N \rangle \langle \bar{p}_T \rangle
          }
      }{
          \left(\frac{\sigma_N}{\langle N \rangle}\right)
      }
   \right] \frac{p_T}{\langle \bar{p}_T \rangle}
\mbox{\hspace*{0.3cm} and \hspace*{0.3cm}}
    v_0^{(2)}(p_T)  \simeq 
    - \frac{3}{2} \frac{\sigma_{p_T}}{\langle \bar{p}_T \rangle} \left(
        1 - \frac{4}{3}\frac{p_T}{\langle \bar{p}_T \rangle}
        \right).
\label{eq:pptoy}        
\end{equation}

Using values obtained from our hydro simulation for $\sigma_N/\langle N \rangle$, $\sigma_{p_T}/\langle \bar{p}_T \rangle$, $\langle \delta N \delta\bar p_T \rangle/   (    \langle N \rangle \langle \bar{p}_T \rangle)$, we can evaluate the analytical model eigenvectors above and compare with the hydrodynamic results in figure \ref{fig:fig1QM}. There is a good agreement and this allows us to extract some information:
    \begin{itemize}
    \item      $v_0^{(1)}(p_T)  \simeq \sigma_N/\langle N \rangle$ is in agreement with CMS data,  so multiplicity fluctuations are correct in our hydrodynamic simulation.
    \item  The $v_0^{(2)}(p_T)$ crossing of the horizontal axis is almost independent of centrality, in agreement with CMS data. This little dependence on centrality arises from the fact that $p_T$ is also almost independent of centrality.
    \item  $v_0^{(2)}(p_T)\propto \sigma_{p_T}/\langle \bar{p}_T \rangle$, i.e., the physical meaning of the second component is connected with the $p_T$ fluctuations (see also \cite{Mazeliauskas:2015efa}).
    \item   The anomalous $v_0^{(1)}(p_T)$ increase in our (and we expect others) hydrodynamic simulation comes from the fact that the $N-p_T$ covariance, and  so the $p_T$ fluctuations are too large.
\end{itemize}

\begin{figure*}
  \centering
\includegraphics[width=\textwidth]{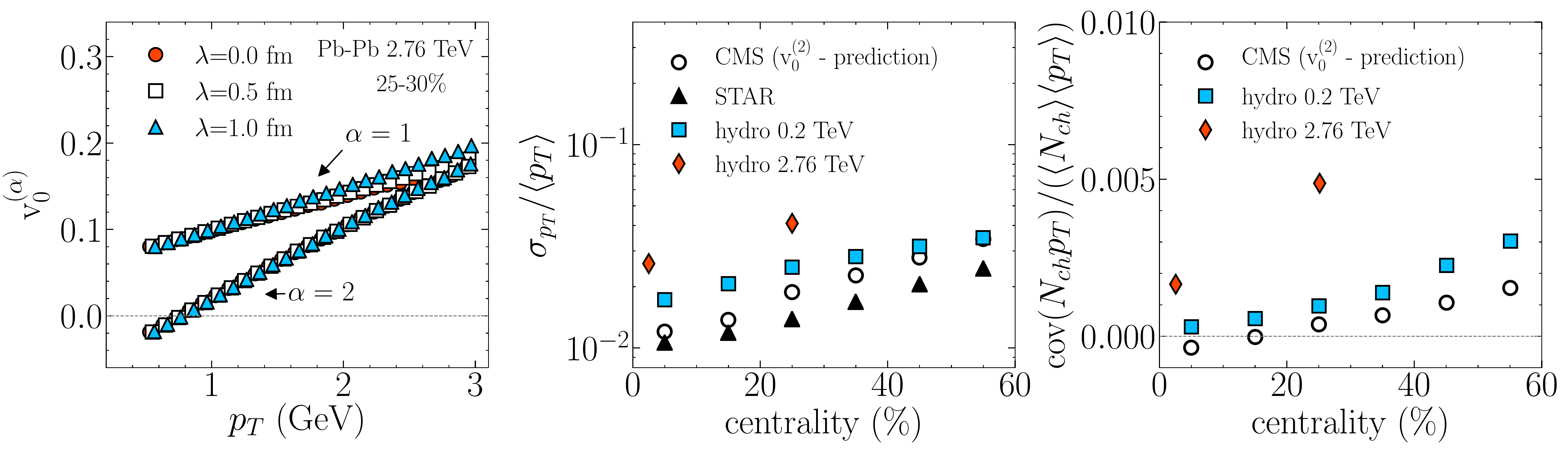} 
\caption{Left: Effect of increasing initial granularity size (corresponds to  increasing $\lambda $ \cite{Gardim:2017ruc}). Center:
  Comparison of   $p_T$ fluctuations from hydrodynamic simulation at RHIC and LHC  with data from STAR~\cite{Adams:2005ka} and an
  estimate from CMS data (see text).   Right:  $N-p_T$ covariance from hydrodynamic simulation at LHC and prediction using CMS data (see text).
}
\label{fig:fig2QM} 
\end{figure*}

We also checked to what extent the size of initial inhomogeneities in the system affects the principal components, smoothing out these inhomogeneities with minimal effect on global properties, as done in Ref. \cite{Gardim:2017ruc}. The left plot of figure~\ref{fig:fig2QM} shows the leading and subleading components have small dependency on the granularity size of the initial condition, parameterized by $\lambda$. A large value of $\lambda$ means that the granularity was smoothed. 

Hydro $p_T$ fluctuations can be compared to data from STAR~\cite{Adams:2005ka}, as well as an estimate from CMS, using the subleading component from the analytical model eq.~(\ref{eq:pptoy}) (center plot of figure~\ref{fig:fig2QM}). At the same time, we can also predict the covariance between $N$ and $p_T$ using CMS data. One can see the covariance is zero for central collisions and increases for peripheral collisions, right panel figure~\ref{fig:fig2QM}. This pattern is the same coming from hydrodynamics.
             
\section{Conclusion}

We compared a Principal Component Analysis for a hydrodynamic simulation with CMS data. While hydro results are reasonable for $n=2$ and 3 \cite{Gardim:2019iah}, for $n=0$ the scaled leading component rises with $p_T$ while data are flat for 2.76 TeV. At 200 GeV, hydrodynamics $v^{(1)}_0$ is flatter than LHC energy. We found that this is due to too large  $N-p_T$ covariance and  $p_T$  variance in the simulation. A natural question is why are these quantities too large? At a given centrality, a smaller size implies a larger density and temperature, hence  larger gradients and mean transverse momentum.
A simple argument allows to relate  $p_T$ fluctuations and  initial size fluctuations \cite{Ollitrault:1991xx,Bozek:2012fw,Gardim:2019brr}: $\sigma_{p_T}/\langle \bar{p}_T \rangle \propto \sigma_r/\langle \bar{r} \rangle$. 

So the origin of the anomalous $v_0^{(1)}(p_T)$ increase is the  too large size of the initial size fluctuations (in our case generated by NeXus). We checked that modifying the  size of the initial hot spots (using the method in \cite{Gardim:2017ruc}) has no  substantial effect as shown in figure~\ref{fig:fig2QM}. In fact there are many models of initial conditions that lead to too large $p_T$ fluctuations \cite{Bozek:2012fw,Broniowski:2009fm}. Since the principal components are sensitive not just to $N$ and $p_T$ fluctuations but their covariance, we expect that the n=0 PCA will provide a stringent test on initial granularity.

\section*{Acknowledgments}
This research was funded by 
CNPq (Conselho Nacional de Desenvolvimento Cient\'{\i}fico) grants 310141/2016-8 (FG),  205369/2018-9 and 312932/2018-9 (FGG), 
FAPEMIG (Funda\c{c}\~ao de Amparo\`a Pesquisa do Estado de Minas Gerais) grant number APQ-02107-16 (FGG), 
FAPESP (Funda\c{c}\~ao de Amparo \`a Pesquisa do Estado
de S\~ao Paulo) grants number  2018/18075-0 (FG), 2018/24720-6 (FG, PI, ML), 2016/24029-6 and 2017/05685-2 (ML),
INCT-FNA grant number 464898/2014-5 (FG, FGG, PI, ML),
Instituto de F\'{\i}sica-USP through graduate student grants from CNPq and CAPES (Coordena\c{c}\~ao de Aperfei\c{c}oamento de Pessoal de N\'{\i}vel Superior) (PI),
USP-COFECUB grant Uc Ph 160-16, 2015/13 (FG, ML, PI, JYO).

\bibliographystyle{elsarticle-num}

\begin{thebibliography}{99}
\bibitem{Bhalerao:2014mua} 
R.~S.~Bhalerao, J.~Y.~Ollitrault, S.~Pal and D.~Teaney,
Phys.\ Rev.\ Lett.\ {\bf 114}, no. 15, 152301 (2015).
\bibitem{Sirunyan:2017gyb} 
A.~M.~Sirunyan {\it et al.} [CMS Collaboration],
Phys.\ Rev.\ C {\bf 96}, no. 6, 064902 (2017).
\bibitem{Mazeliauskas:2015vea} 
A.~Mazeliauskas and D.~Teaney,
Phys.\ Rev.\ C {\bf 91}, no. 4, 044902 (2015).
\bibitem{Mazeliauskas:2015efa} 
A.~Mazeliauskas and D.~Teaney,
Phys.\ Rev.\ C {\bf 93}, no. 2, 024913 (2016).
\bibitem{Liu:2019jxg}
Z.~Liu, W.~Zhao and H.~Song, Eur. Phys. J. {\bf C  79}, no.10,  870 (2019) .
\bibitem{Bozek:2017thv} 
P.~Bozek,
Phys.\ Rev.\ C {\bf 97}, no. 3, 034905 (2018).
\bibitem{Hippert:2019swu}
M.~Hippert, D.~D.~Chinelatto, M.~Luzum, J.~Noronha, T.~S.~Silva and J.~ Takahashi,
[arXiv:1906.08915 [nucl-th]].
 \bibitem{Gardim:2019iah}
 F.~G.~Gardim, F.~Grassi, P.~Ishida, M.~Luzum, J.Y.~Ollitrault, 
 Phys. Rev. {\bf C100}, n.5, 054905 (2019)
[arXiv:1906.03045 [nucl-th]].
\bibitem{Matt}
M. ~Luzum, private communication about MUSIC code with Trento initial conditions.
\bibitem{Adams:2005ka}
J.~Adams {\it et al.} [STAR Collaboration], Phys.\ Rev.\ C {\bf 72}, 044902 (2005).
\bibitem{Gardim:2017ruc}
F.~G.~Gardim, F.~Grassi, P.~Ishida, M.~Luzum, P.~S.~Magalhães and J.~Noronha-Hostler,
Phys.\ Rev.\ C \textbf{97}, no.6, 064919 (2018)
[arXiv:1712.03912 [nucl-th]].
\bibitem{Ollitrault:1991xx}
J.-Y.~Ollitrault, Phys. Lett. {\bf B273} 32 (1991).
\bibitem{Bozek:2012fw}
P.~Bozek and W.~Broniowski, Phys.\ Rev.\ C {\bf 85}, 044910 (2012).
\bibitem{Gardim:2019brr}
F.~G.~Gardim, G.~Giacalone and J.~Ollitrault,
[arXiv:1909.11609 [nucl-th]].
\bibitem{Broniowski:2009fm}
W.~Broniowski, M.~Chojnacki and L.~Obara, Phys.\ Rev.\ C {\bf 80}, 051902 (2009).
 
\end{thebibliography}

\end{document}